\documentclass[jcp,amsmath,amssymb,amsfonts,superscriptaddress,preprint]{revtex4-1}

\usepackage{graphicx}
\usepackage{natbib}

\usepackage{verbatim}

\begin{document}

\title{\textbf{Plasmonic opals: observation of a collective molecular exciton mode beyond the strong coupling}}

\author{Pierre Fauche}
\affiliation{CNRS, University of Bordeaux, CRPP, UPR8641, 115 av. Schweitzer, 33600 Pessac, France}
\author{Christian Gebhardt}
\affiliation{CNRS, University of Bordeaux, CRPP, UPR8641, 115 av. Schweitzer, 33600 Pessac, France}
\author{Maxim Sukharev}
\email{maxim.sukharev@asu.edu}
\affiliation{Arizona State University, Mesa AZ 85212, USA}
\author{Renaud A. L. Vallee}
\email{vallee@crpp-bordeaux.cnrs.fr}
\affiliation{CNRS, University of Bordeaux, CRPP, UPR8641, 115 av. Schweitzer, 33600 Pessac, France}

\begin{abstract}
Achieving and controlling strong light-matter interactions in many-body systems is of paramount importance both for fundamental understanding and potential applications. In this paper we demonstrate both experimentally and theoretically how to manipulate strong coupling between the Bragg-plasmon mode supported by a  organo-metallic array and molecular excitons in the form of J-aggregates dispersed on the hybrid structure. We observe experimentally the transition from a conventional strong coupling regime exhibiting the usual upper and lower polaritonic branches to a more complex regime, where a third nondispersive mode is seen, as the concentration of J-aggregates is increased. The numerical simulations confirm the presence of the third resonance. We attribute its physical nature to collective molecule-molecule interactions leading to the collective electromagnetic response. A simple analytical model is proposed to explain the physics of the third mode. The nonlinear dependence on molecular parameters followed from the model are confirmed in a set of rigorous numerical studies. It is shown that at the energy of the collective mode molecules oscillate completely out of phase with the incident radiation acting as an effective thin metal layer.
\end{abstract}

\maketitle

\section{Introduction}
The field of nanoplasmonics\cite{Stockman:11} is rapidly expanding to various research fields including materials science and chemistry due to tremendous capabilities offered by modern nanofabrication techniques.\cite{ADMA:ADMA200700678} Outstanding control is achieved in patterning metal surfaces.\cite{Gramotnev:2014aa} Such a control provides a wider accessibility to evanescent electromagnetic modes to manipulate individual atoms and molecules.\cite{Pelton:2015aa,Lodahl:2015aa} Being able to couple molecules to cavity modes in the strong coupling regime, i.e. when the coupling strength surpasses all damping rates, is appealing due to various applications in molecular and material science by changing and allowing one to control the optical properties of the emitters or absorbers.

Recently, the strong optical coupling of plasmonic and excitonic systems has been achieved for different plasmonic structures, involving both localized~\cite{Bellessa2009,Balci2013,Zengin2013,Hakami2014} and propagating~\cite{Bellessa2004,Hakala2009,Balci2012} surface plasmon-polaritons. Hybrid exciton-plasmon modes formed due to strong coupling between molecules and plasmons have been observed as upper and lower polariton branches.\cite{Torma2015a} The separation between the branches at zero-detuning referred to as a Rabi splitting determines how strong two subsystems are coupled and how quickly the energy is transferred between them. The research in exciton-plasmon systems has its origins in the physics of semiconductor microcavities, where a typical Rabi splitting can easily achieve tens of meV.\cite{Khitrova:2006aa} Plasmon-sustaining nanomaterials offer significantly smaller volume of a resonant mode (in this case surface plasmon-polaritons) and thus higher local fields and in turn greater Rabi splittings well surpassing $100$ meV mark.\cite{doi:10.1021/nl4014887} 

In this paper, we examine the light-matter interaction in the strong coupling limit by combining the Bragg-plasmon (BP) mode of a hybrid organo-metallic colloidal array with excitonic J-aggregates dispersed on top of the grating. In order to control and maximize the Rabi splitting energy, the spatial and spectral confinements of the electric field in immediate proximity of the curved metallic surface match the resonance of J-aggregates contained in a polymer matrix dispersed on the top. Furthermore, the concentration of the J-aggregates is varied such as to control the transition dipole strength of the excitonic system. We determine the dispersion relations of our systems for the various concentrations by varying the incident wavevector (angular tuning) in reflection experiments. Very interestingly, as the excitonic system reaches a characteristic concentration, we observe the transition from a traditional strong coupling regime with two polaritonic branches to a new regime of high molecular concentrations, where a third branch, nearly dispersionless, appears in the reflection spectra, starting from a characteristic incident angle.

To explain the unusual experimental results we perform rigorous numerical simulations of the exciton-plasmon opal arrays. We also propose and test a simple analytical model that explains the physical origin of the third mode and links its appearance in reflection and transmission spectra to a collective resonance due to strong molecule-molecule interactions. We also apply the proposed theoretical model to other hybrid systems finding that the observed phenomenon is general and should be observed in other systems thus confirming and expanding the earlier theoretical results.\cite{PhysRevLett.109.073002}

\section{Experimental results}
\begin{figure}
\centering
\includegraphics[width=12cm]{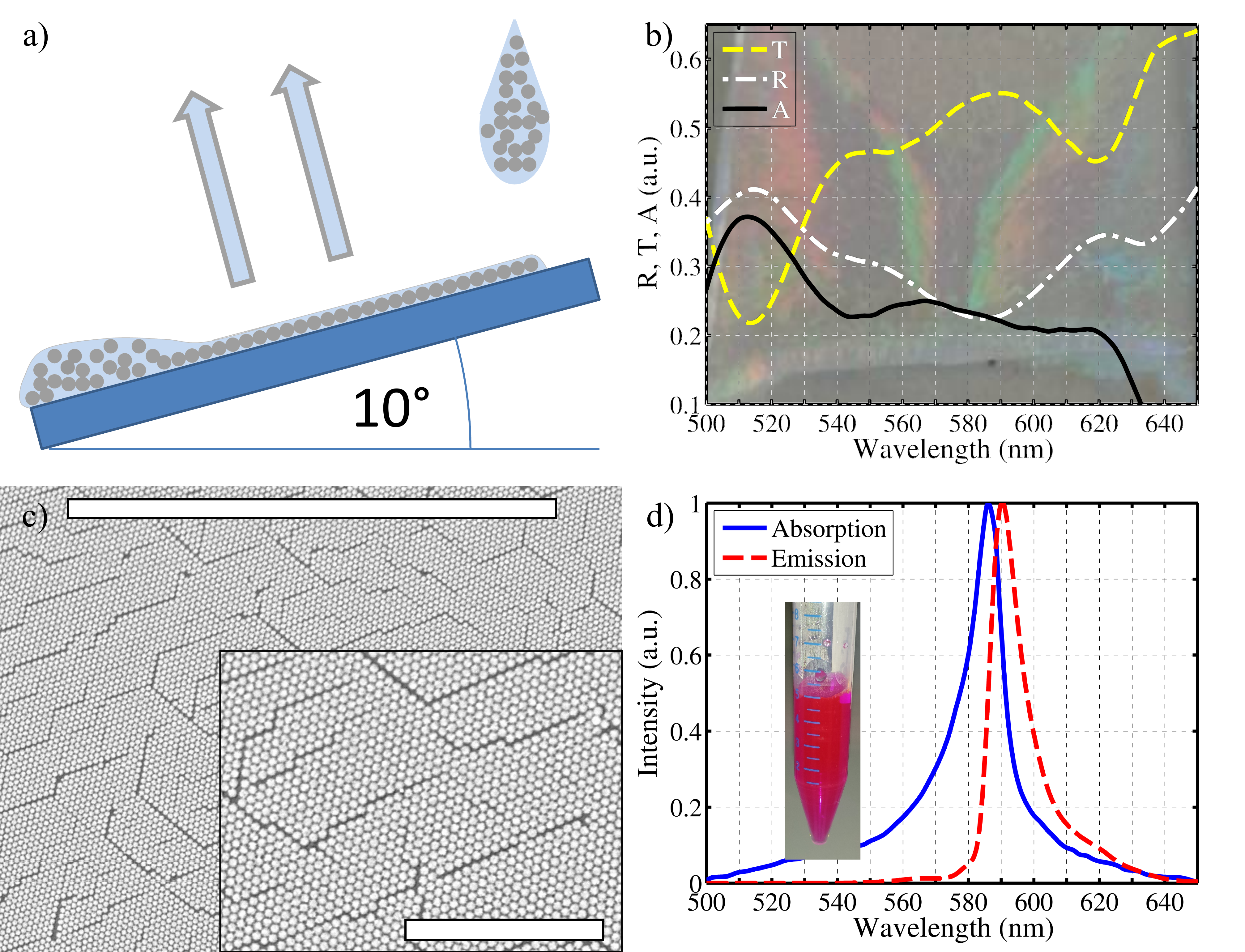} 
\caption{Panel (a) shows a schematics of the convective self-assembly of an hexagonal mono-layer of polystyrene (PS) beads onto a glass substrate. Panel (b) shows transmission, $T$, reflection, $R$, and absorption, $A$ spectra for the 2D plasmonic opal with beads of diameter $D = 478$ nm. The background figure shows the visible iridescence of the structure. Panel (c) shows SEM images at low magnification, with a scale bar of 50 $\mu$ m. Panel (d) shows absorption (solid blue line) and emission (dashed red line) spectra of TDBC in NaOH solution (50 $\mu$mol/L), a photography of this TDBC solution is shown in inset.}
\label{fig:material}
\end{figure}

Two-dimensional plasmonic opals were manufactured by Nano-Sphere Lithography. Briefly, hexagonal close packed mono-layers of polystyrene (PS) micro-spheres (Polysciences, diameters $D = 478$ nm, 2.5\% solids (w/v) aqueous suspensions) were deposited on UV/ozone-cleaned (30 min) 5 mm thick glass plates with a size of $2\times 2$ cm$^{2}$ at a tilt angle of approximately 10$^o$. Figure ~\ref{fig:material} a) shows a schematic setup of this process performed under controlled conditions of temperature and humidity (T $=293$ K, H $=65$\%). In a second step,~\cite{Ungureanu2013,Fauche2014} a 50 nm thick silver layer was deposited on top of these mono-layers. Finally, the organic excitonic system (TDBC cyanine dyes known to form J-aggregates with a large transition dipole moment) was formed by spin-coating a highly-concentrated solution of the dye molecules dissolved ($5$ mg/mL, $10$ mg/mL) in a $90:10$ water:ethanol mixture, further bathed for $2$ min in a $2$\% PDAC solution, onto the silver corrugated surface.

The good quality and periodicity of these 2D plasmonic opals were ascertained optically by direct visualization of their visible iridescence (Figure ~\ref{fig:material}b), as well as by their transmission (T) and reflection (R) spectra recorded at normal incidence through a 10x objective (ZEISS Ultrafluar 10x/0.2) on a  micro - spectrophotometer (20/20 V CRAIC Technologies). Figure ~\ref{fig:material}d also shows the absorption (A) spectrum, obtained by using the usual formula $A(\lambda)=1-T(\lambda)-R(\lambda)$. The optical transmission of a metal coated colloidal monolayer measured at normal incidence is known to exhibit a transmission maximum.\cite{Landstrom2006a} This maximum is due to the coupling of the incident light to a  surface plasmon-polariton mode via Bragg scattering on the 2D lattice. Fig.~\ref{fig:material}b shows such a transmission maximum at $592$ nm, which is expected for a structure based on $D=478$ nm diameter PS beads. In agreement with the literature,~\cite{Romanov2011} the Bragg plasmon extraordinary transmission is associated with a reflectance minimum slightly shifted to shorter wavelength of $587$ nm. The samples are characterized by Scanning Electron Microscopy (Hitachi TM-1000). Figure \ref{fig:material}c shows the metal-coated dielectric crystalline plane over an area of 50 $\mu$m x 50 $\mu$m. From the top left of the figure to the bottom right, a single crystalline orientation is found. If some grooves are observed between two adjacent lines of beads (either caused by attractive capillary forces acting at a late stage of drying or by the presence of a tiny or a big bead slightly disturbing the arrangement), they do not reach the state of dislocations, or grain boundaries, as there are no orientation changes between adjacent domains.

The measured absorption and fluorescence spectra of TDBC J-aggregates in NaOH solution, shown in Figure~\ref{fig:material}d, were recorded by means of a UV/Vis/NIR spectrophotometer (Lambda 950 PERKIN ELMER) and a spectrofluorometer (JASCO FP-8300), respectively.  From the absorption spectrum of 10 different samples, we determined the TDBC exciton energy and its width at half maximum to be $2.12$ eV and $48.8$ meV, respectively. Atomic Force Microscopy measurements (AFM) confirmed a thickness of 10 nm for the PDAC embedded TDBC film spin-coated on the hybrid structure.

\begin{figure}
\centering
\includegraphics[width=12cm]{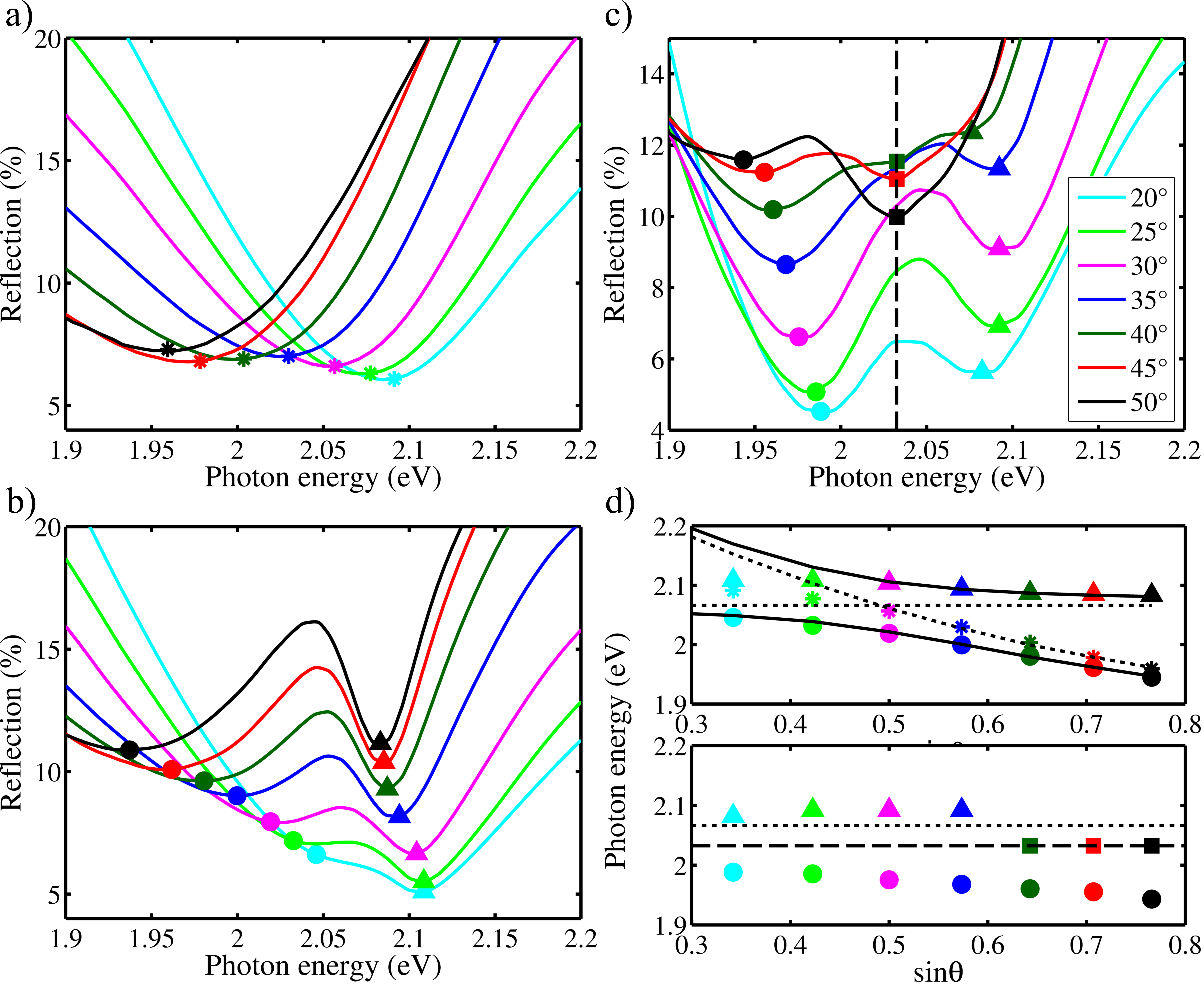} 
\caption{Experimental reflection spectra of a plasmonic opal as a function of incidence angle for the following structures: (panel (a)) a bare opal array without molecules, (panel (b)) covered by TDBC molecules at low concentration in the PDAC film, and (panel (c)) covered by TDBC molecules at high concentration in the PDAC film. The top of panel (d) shows dispersion relations of the experimentally observed i) lower (shown as circles) and ii) upper (shown as triangles) polaritonic branches extracted from spectra in case of a low concentration of TDBC molecules together with iii) the bare Bragg plasmon mode (shown as stars). Their corresponding calculated branches are also represented as solid and dashed line, respectively. The non dispersive dashed line corresponds to the transition energy of the TDBC J-aggregates. The bottom of panel (d) shows dispersion relations of the experimentally obtained i) lower (shown as circles) and ii) upper (shown as triangles) polaritonic branches obtained in case of a high concentration of TDBC molecules together with iii) the third mode (squares). The short and long dashed lines correspond to the transition energy of the TDBC J-aggregates and the calculated dispersion-less relation of the third mode.}
\label{fig:results}
\end{figure}

The UV-visible reflection spectra of the 2D plasmonic opal measured at different incident angles are shown in Figure ~\ref{fig:results}a. These spectra were obtained by collecting the reflectance from a large area ($3\times 3$ mm$^{2}$). A dispersive behavior of the Bragg plasmon mode is clearly noticed with the evolution of the reflectance minimum (indicated as stars in the figure) as a function of incidence angle. The related dispersion curve is shown in Fig. ~\ref{fig:results}d. Assuming that the incident field couples to a mode with an effective wave vector $\sqrt{\epsilon_\text{eff}}k$ set at an azimuthal angle $\phi$ with respect to the plane of incidence, the position of the reflection dip can be estimated using the formula for the first-order diffraction ~\cite{Landstrom2006} 

\begin{equation}
\lambda=\frac{\sqrt{3}D}{2}(\sqrt{\epsilon_\text{eff}-\sin^{2}\theta \sin^{2}\phi}-\sin\theta \cos\phi),
\label{eq:BP}
\end{equation}
where $D = 478$ nm is the diameter of PS spheres, estimated from SEM micrographs. $\epsilon_\text{eff}$ is the effective permittivity expressed by the mixing rule\cite{Ding2013} $\frac{1}{\epsilon_\text{eff}(\lambda)}=\frac{1}{\epsilon_\text{Ag}(\lambda)}+\frac{1}{\epsilon_\text{dielectric}}$in which $\epsilon_\text{dielectric}$ is defined on a filling factor basis $\sqrt{\epsilon_\text{dielectric}}=f_{PS}n_{PS}+(1-f_{PS})$.\cite{Landstrom2005} The filling factor was deduced from the position of the Bragg resonance for a metal non-coated PS hexagonal lattice: $f_{PS}=0.6$. The refractive index of the PS spheres we used was $n_{PS}=1.57$.\cite{Ma2003} The dielectric permittivity of silver was taken from the model proposed by Johnson and Christy.\cite{Johnson1972}

As the reflection spectra are collected over large area ($3\times 3$ mm$^{2})$, several domain orientations contribute to the measurement process. The measured reflectance minima thus result from an averaging which depends on both azimuthal angles $\phi$ and angles of incidence $\theta$. Since it has been reported that $s-$ and $p-$polarized light preferentially couple to angles $\phi=120^{\circ} \pm 3^{\circ}$ and $\phi=79^{\circ} \pm 12^{\circ} $, respectively,\cite{Landstrom2006} we provide the best dispersion line based on our results in Fig. ~\ref{fig:results}d (dashed line). Clearly, a small discrepancy between our experimental results and the first-order diffraction law occurs at small $\theta$ angles. This results from the fact that we take only into account one azimuthal angle in our fit, which clearly underestimates the multi domain orientations seen by light at small incidence angle.  
 
We performed complete characterization of the bare plasmonic opals and found experimental results in good agreement with already published data.\cite{Landstrom2006,Romanov2011,Ding2013} We now turn our attention to the interaction between the Bragg plasmon mode and the TDBC J-aggregates' excitons. Two types of films are here considered, depending on the concentration of TDBC molecules involved: the low (5 mg/mL) and high (10 mg/mL) TDBC film's concentrations. 

The UV-visible reflection spectra of the 2D plasmonic opals covered with a small TDBC film's concentration are shown in Fig. ~\ref{fig:results}b as a function of angle ($\theta$). Owing to the interaction between the large transition dipole moment of the TDBC J-aggregates and the Bragg plasmon enhanced field, a clear Rabi splitting is observed with a red shift of the upper and lower polaritonic branches as the incidence angle increases, as best observed in Fig. ~\ref{fig:results}d (colored symbols). Following the conventional coupled oscillators model\cite{Agranovich2003,Bellessa2004}, we can describe the hybrid states form due to the exciton-plasmon interaction
 \begin{equation}
E_{U,L}(k)=\frac{1}{2}\left(E_{\text{BP}}(k)+E_0\right) \pm \sqrt{\Delta+\frac{1}{4}\left(E_{\text{BP}}(k)-E_0\right)^2},
\label{eq:SC}
\end{equation}
where $k$ is the in-plane wavevector, $E_0$ denotes the molecule transition energy, $E_{\text{BP}}$ is the energy of the Bragg plasmon mode, $2\Delta$ is the Rabi splitting. Fig. ~\ref{fig:results}d (top) shows that the TDBC exciton crosses the Bragg plasmon dispersion line. As a consequence, the anti-crossing of the upper and lower polaritonic branches result from the strong coupling between the Bragg-plasmon mode and the TDBC J-aggregates' exciton. The calculated hybrid states are shown as black solid lines. As the spin-coating of a PDAC embedded TDBC J-aggregates film induces a slight increase of the effective refractive index as compared to the bare Bragg-plasmon mode, we adjusted the calculated polaritonic branches on the measurement data owing to a slight change of the effective refractive index. The Rabi splitting energy $2\Delta$ has been estimated at $85\pm 5$ meV.

Very interestingly, while increasing the J-aggregates' concentration, we clearly observed, in a reproducible way, a significantly different trend with respect to the one reported above, for which the observed Rabi splitting exhibits a red shift of the lower polaritonic branch and a blue shift of the higher polaritonic branch (Fig. ~\ref{fig:results}c) as the incidence angle increases. Furthermore, from an incidence angle of about 35$^o$, a new mode emerges as a dip in the reflection spectra, at $2.03$ eV. The evolution of the new mode as a function of the incidence angle results in a new, nearly dispersionless, branch as seen in Fig. ~\ref{fig:results}d (bottom). This feature is very unique for the kind of a structure we are investigating. It points to the manifestation of a collective many-body effect originating from the large concentration of excitons interacting with the Bragg-plasmon mode of the structure.  

In order to get further insight into the nature of the new electromagnetic mode, we performed a series of rigorous numerical simulations making use of fully vectorial model based on coupled Maxwell-Bloch equations. We also propose a simple analytical model capable of explaining the physics of the third resonance linking it to the collective molecular exciton mode predicted theoretically.\cite{PhysRevA.84.043802} 

\section{Theoretical model}
We model the optical response of opal arrays by numerically integrating coupled Maxwell-Bloch equations in three dimensions. The dielectric function of silver is modeled using conventional Drude function with the parameters relevant to the spectral range of interest.\cite{Blake:2015aa} The molecular system is modeled via numerical integration of rate equations describing interacting two-level emitters driven by a local electric field.\cite{Sukharev:2016aa} The resulting system of equations is solved self-consistently with no further approximations. Due to memory limitations and numerical convergence issues at high molecular concentrations we were not able to simulate opal arrays with a period greater than $360$ nm. Our simulations were also limited to the normal incidence case. Thus our goal is to achieve a qualitative understanding of the experimental observations. In order to obtain proper dependence of the energies of hybrid modes on the exciton-plasmon coupling we vary periodicity of the array rather than the angle of incidence. In most cases this leads to expected behavior supported by experimental observations. We choose the period of $320$ nm as our primary target to couple Bragg plasmons to molecules. Simulations of reflection and transmission spectra of a bare opal array at this period reveal a set of resonances that include a localized surface plasmon-polariton mode at $2.3$ eV, the Bragg plasmon at $2.94$ eV, and higher order Bragg modes at higher energies. The analytical model\cite{Fauche2014} predicts the energy of the Bragg plasmon for this period to be near $2.87$ eV, which matches well our simulations. We note that the complexity of the local EM field distribution increases with increasing energy.
\begin{figure}
\centering
\includegraphics[width=12cm]{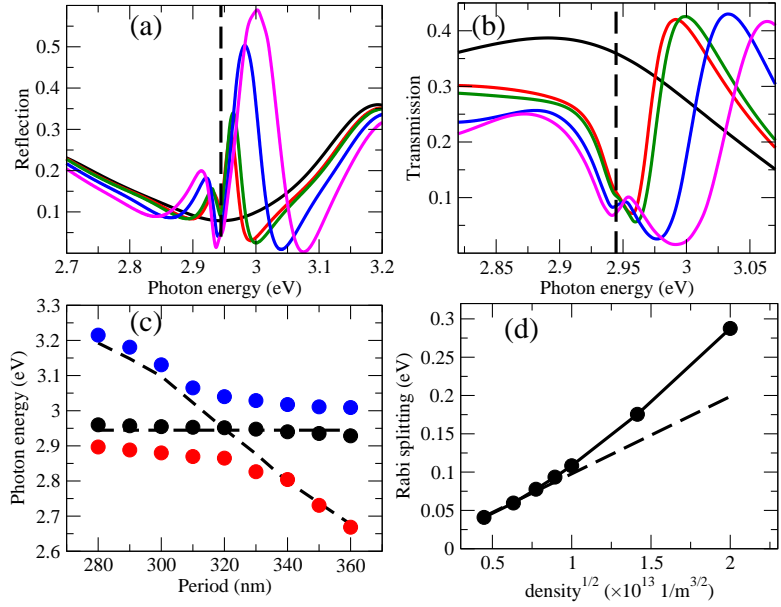}
\caption{Results of numerical modeling I. Panels (a) and (b) show reflection and transmission as functions of the incident photon energy for the hybrid system comprising polystyrene spheres with a period of $320$ nm covered by a silver film with a thickness of $50$ nm and molecular layer with a thickness of $20$ nm. Black lines show results for the bare opal array, red lines are for the the hybrid system at the molecular concentration of $8\times10^{25}$ m$^{-3}$, green lines are for $10^{26}$ m$^{-3}$, blue lines are for $2\times10^{26}$ m$^{-3}$, and magenta lines are for $3\times10^{26}$ m$^{-3}$. Vertical dashed line in both panels indicates the molecular transition energy corresponding to the Bragg-plasmon mode for the array with a $320$ nm period ($2.94$ eV). Panel (c) shows energies of the upper (blue circles) and lower (red circles) polaritons as functions of the period of the opal array at the molecular concentration of $2\times10^{26}$ m$^{-3}$. The black circles indicate the position of the third mode extracted from transmission spectra. Horizontal dashed line shows the molecular transition energy. The curved dashed line shows the dependence of the Bragg-plasmon mode of the bare opal array on the period. Panel (d) shows the Rabi splitting as a function of the $\sqrt{n}$, where $n$ is the number density of molecules. The straight dashed line shows the expected $\sqrt{n}$ functional dependence. Other molecular parameters are: the transition dipole moment is $d_0=10$ Debye, the radiationless lifetime of the excited state is $1$ ps, the pure dephasing time is $730$ fs}
\label{fig:simulations1}
\end{figure}

Fig. \ref{fig:simulations1} presents simulated spectra of plasmonic opals with a $20$ nm thin molecular layer on top. At low molecular concentrations two hybrid states, namely upper and lower polaritons are observed as minima in reflection (panel (a)) and maxima in transmission (panel (b)). When the density increases the third resonance is seen near the molecular transition energy. It is interesting to note that it first appears in reflection and then, at higher concentrations, pops up in transmission with a blue-shifted energy with respect to the molecular transition. Its position further moves to higher energies with increasing concentration while the resonance becomes broader. The third mode is nearly dispersionless. It exhibits small deviations from the molecular transition line due to repelling by the lower polariton at shorter periods and by the upper polariton at longer periods, as seen in Fig. ~\ref{fig:simulations1}c. 

\begin{figure}
\centering
\includegraphics[width=9cm]{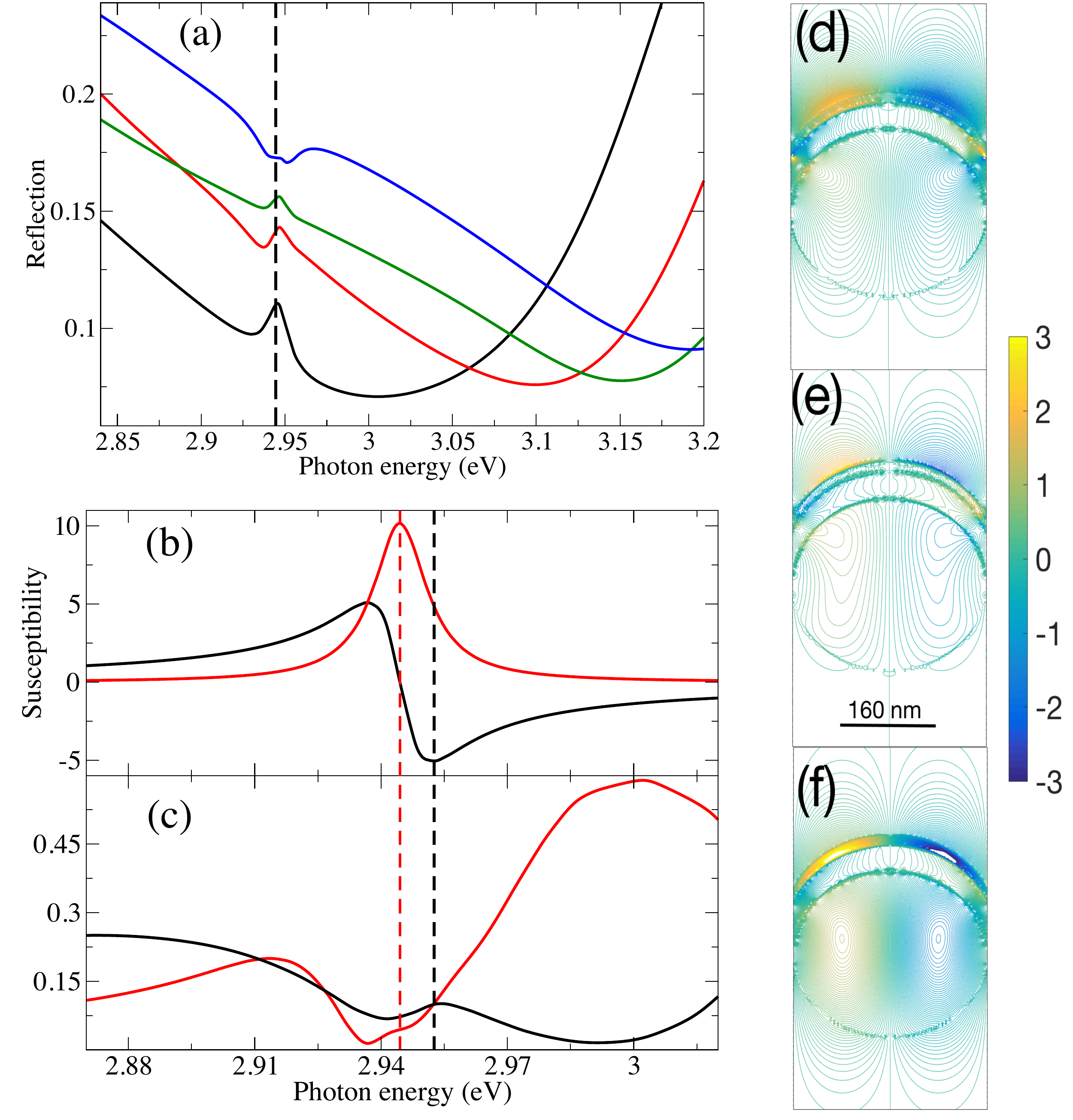}
\caption{Results of numerical modeling II. Panel (a) shows the reflection spectra calculated at different periods of the opal array. The molecular density is $10^{25}$ m$^{-3}$. Black line shows results for the period of $310$ nm, red line is for $300$ nm, green line is for $290$ nm, and blue line shows data for $280$ nm. Panel (b) shows the molecular electric susceptibility, $\chi_M$, calculated at the molecular density of $3\times10^{26}$ m$^{-3}$. The black line shows the real part of $\chi_M$, the red line shows the imaginary part of $\chi_M$. The vertical red dashed line indicates the molecular transition energy (corresponding to the maximum of Im$(\chi_M)$). The vertical black dashed line shows the energy corresponding to the minimum of Re$(\chi_M)$. Panel (c) shows transmission (black line) and reflection (red line) for the period of $320$ nm and molecular density of $3\times10^{26}$ m$^{-3}$. Panels (d) - (f) show spatial distributions of the longitudinal component of electric field, $E_z$, normalized with respect to the incident field amplitude. The incident field excites the system along $x$ (horizontal axis in each panel) axis propagating from top to bottom of the panels. Panel (d) shows $E_z$ for the lower polariton mode calculated at the energy $2.86$ eV. Panel (e) shows $E_z$ at the energy of the third mode, namely $2.94$ eV. Panel (f) shows $E_z$ at the energy of the upper polariton $3.04$ eV. The plane of calculations is at $y=0$. Other parameters of the simulations are: the diameter of polystyrene spheres is $320$ nm, the molecular concentration is $2\times10^{26}$ m$^{-3}$, the molecular transition dipole moment is $10$ Debye, the radiationless lifetime of the excited state is $1$ ps, and the pure dephasing time is $730$ fs}
\label{fig:simulations2}
\end{figure}

The advantage of our model that relies on Maxwell-Bloch equations is that it captures many-body effects resulting from strong molecule-molecule interactions due to a high local electric field and high molecular densities. It is thus informative to test the range of densities for which the conventional model of two coupled oscillators \eqref{eq:SC} could be applied. The model \eqref{eq:SC} predicts that the Rabi splitting should scale as a $\sqrt{n}$, where $n$ is the number density of molecules. We numerically extract the Rabi splitting from reflection spectra at different densities (note that similar functional dependance is obtained for the Rabi splitting evaluated using transmission instead of reflection). The results are shown in Fig. ~\ref{fig:simulations1}d. Indeed the Rabi splitting scales as a square root of molecular concentration at low densities as expected but begins to deviate noticeably from predicted dependence at concentrations above $6\times10^{25}$ m$^{-3}$. This suggests that molecule-molecule interactions may play a significant role in formation of hybrid states and may, in principle, be responsible for the appearance of the third mode as well.

Our qualitative attempt to reproduce experimental spectra (Fig.  \ref{fig:results}c) is presented in Fig. \ref{fig:simulations2}a. A small distortion in the reflection near the molecular resonance is observed. This knee structure indicates the presence of the third mode, which becomes more pronounced at higher molecular densities. With the period decreasing (corresponding to the increase of the in-plane $k$-vector, i.e. the increase of the angle of incidence) we observe a clear development of the third mode, which is blue-shifted with respect to the molecule resonance. It should be noted that the upper polariton at the period of $280$ nm is located near $3.2$ eV and has no influence on the third mode. 

The physics of the third mode is readily understood through comparing Figs. \ref{fig:simulations2}b and c. Panel (b) shows the real and imaginary parts of the molecular susceptibility calculated at $3\times10^{26}$ m$^{-3}$. Plotted below this panel are transmission and reflection evaluated at the same density. One can see that both transmission and reflection have minima at the molecular transition energy (red dashed line) while the transmission exhibits the resonance at the energy corresponding to the minimum of Re$(\chi_M)$. The molecules thus oscillate out-of-phase with respect to the incident field essentially resulting in an effective \textit{metallic} layer covering the silver film. The resonance observed at high molecular concentrations is thus due to the formation of a new Bragg-plasmon type mode with molecules acting as a metallic surface. We note here that the observed \textit{metallic} behavior of molecules is yet another indicator of the collective nature of the third mode.

To further scrutinize the origin of the third mode we examine steady-state distributions of the longitudinal component of electric field for three resonances. The results of simulations are shown in Fig. \ref{fig:simulations2}d-f. Firstly, we note that upper and lower polaritons are characterized by a substantial field inside the molecular layer as seen in panels (d) and (f). Secondly, both these modes show a clear dipolar pattern with dipoles formed inside molecular layer, metal (with opposite signs), and polystyrene spheres. The obvious distinction between these two modes is how the field is distributed in and around the molecular layer. In case of the lower polariton mode (Fig. ~\ref{fig:simulations2}d) such a distribution shows the smooth extension of the field outside of the molecular layer. Oppositely to the latter the upper polariton (Fig. ~\ref{fig:simulations2}f) has the field significantly localized inside the molecular film. The third mode (Fig. ~\ref{fig:simulations2}e) is clearly distinct from both hybrid states as it has low electromagnetic field inside the molecular film but highly localized evanescent field on the outer surface of the molecular subsystem. The field inside molecular layer and silver is nearly uniform. 

In order to explain the very presence of three resonant energies in spectra of essentially two coupled systems (the Bragg plasmon mode on the one side and a set of identical molecules on the other), we propose a simple quantum mechanical model expanding the conventional $2\times2$ model to include $N$ identical molecules. This model also elucidates the nature of the observed third mode and predicts the proper dependence of the Rabi splitting on the molecular concentration and the transition dipole. The Hamiltonian of $N$ identical molecules interacting with one another and with the surface plasmon-polariton field reads as
\begin{equation}
\label{Hamiltonian}
\hat{H}=
\left(
\begin{array}{ccccc}
E_0 & C & C &... & \Delta \\
C & E_0 & C & ... & \Delta \\
C & C & E_0 & ... & \Delta \\
... & ... & ... & ... & ... \\
\Delta & \Delta & \Delta & ... & E_{\text{PL}}
\end{array}
\right),
\end{equation}
where we introduce the phenomenological constant $C$ describing the coupling between molecules. We note that this model assumes that molecules are interacting with one another with the same strength. This assumption obviously holds as long as the size of a system is smaller than the corresponding wavelength. In our case the resonant wavelength is of the same order as a PS bead diameter. This means that the model neglects retardation effects. As we will see below, qualitatively, the proposed model predicts the correct behavior of the third resonance while if one needs to describe it quantitatively, the numerical integration of Maxwell-Bloch equations is required.

The Hamiltonian (\ref{Hamiltonian}) can be analytically diagonalized leading to the following set of eigenenergies
\begin{align}
\label{eigenstates}
E_n&=E_0-C, n=1,...,N-2,
\\
E_{N-1,N}&=\frac{1}{2}\left( \left(N-1\right)C+E_0+E_{\text{BP}}\right)\pm\sqrt{\frac{1}{4}\left( \left(N-1\right)C+E_0-E_{\text{BP}}\right)^2+N\Delta^2},
\end{align}
where the first root is $\left(N-2\right)$-degenerate and the other two clearly represent the upper and lower polaritonic branches although notecably altered by molecule-molecule interactions. From the proposed model the Rabi splitting, $\Delta E$, defined as the energy difference between upper and lower polaritons at the zero-detuning condition $E_0=E_{\text{PL}}$ is
\begin{equation}
\label{RabiSplitting}
\Delta E=\sqrt{\left(N-1\right)^2C^2+4N\Delta^2}.
\end{equation}
At low molecular concentrations, as the molecule-molecule interaction energy $C$ is negligible, the Rabi splitting (\ref{RabiSplitting}) scales as a $\sqrt{N}$ as in the conventional model \eqref{eq:SC}. Moreover, the scaling of the Rabi splitting with respect to the molecular transition dipole, $d_0$, is linear since the molecule-plasmon coupling constant $\Delta$ also scales linearly with $d_0$. Our model thus contains the model (\ref{eq:SC}) as a limiting case. 

Let us now examine the behavior of the Rabi splitting at high molecular concentrations. The scaling with molecular density at high $N$ reads as
\begin{equation}
\label{RabiSplittingN}
\Delta E\approx N\sqrt{C^2+4\frac{\Delta^2}{N}}\approx NC.
\end{equation}
The clear deviation of the Rabi splitting from the $\sqrt{N}$ dependence is seen in (\ref{RabiSplittingN}). 

We note that we still have to determine how the coupling constant $C$ scales with molecular parameters. In order to identify the physical nature of $C$ we performed a set of extensive simulations varying the molecular transition dipole moment and molecular density at the resonant conditions corresponding to the period of $320$ nm. From simulations we observe a nearly ideal quadratic dependence of the coupling, $C$, on the dipole moment and a clear linear scaling with the molecular density at intermediate densities. Using a simple argument of the mean inter-particle distance scaling with the number density $\left<R\right>\sim1/N^{1/3}$, we arrive at the following expression for the coupling constant $C$
\begin{equation}
\label{ConstantC}
C\sim Nd_0^2\sim d_0^2/R^3,
\end{equation}
where $R$ is an average molecule-to-molecule distance. The coupling constant evidently has a form of the potential energy of a dipole $d_0$ in the field of another identical dipole thus confirming earlier prediction\cite{PhysRevLett.109.073002} that this mode corresponds to the collective molecule-to-molecule resonance. Moreover this expression combined with (\ref{RabiSplittingN}) results in a clear quadratic dependence of the Rabi splitting on the molecular density. Our analytical model is independent from the geometry suggesting that the results discussed here should in principle be observed in other hybrid systems. The presence of additional resonances in hybrid systems at high molecular concentrations is not limited to periodic systems. The third mode was first observed in numerical simulations in core-shell nanoparticles\cite{PhysRevA.84.043802} and then later on confirmed using different models.\cite{doi:10.1021/ph500032d} The latter presented a geometrical interpretation of this resonance arguing that its physical nature is related to the negative effective permittivity and high absorption of a molecular system.

We note that the collective mode discussed above does not require a presence of the plasmon field. This mode should be observed in molecular clusters at high densities as it was demonstrated elsewhere.\cite{PhysRevA.84.043802} All attributes of the collective nature of this mode pointed out above such as quadratic dependance on the transition dipole were found in excitonic clusters as well.\cite{PhysRevA.84.043802} One can argue that for a given molecular concentration higher exciton-plasmon coupling (i.e. higher values of the Rabi splitting) may in principle lead to an observation of the collective mode as long as the damping at the molecular transition energy is lower than the strength of molecule-molecule interaction.

\section{Conclusion}
In summary, we performed extensive experimental and theoretical studies of the hybrid nanostructure comprising a periodic opal array coupled to a thin molecular layer. It is shown that the system exhibits strong coupling as readily seen in reflection spectra. The observed upper and lower polaritonic branches follow a conventional coupled oscillators model as seen in both experiment and numerical simulations. At high molecular concentrations, a third resonance is observed and is attributed to a collective electromagnetic mode. Rigorous numerical simulations and a simple analytical model shed light onto the physical nature of the new mode. It is shown that such a mode corresponds to the collective molecular exciton resulting from strong molecule-molecule interaction. The dependence of the energy of this mode on various material parameters is confirmed by numerical simulations and further explained using a simple analytical model. It is demonstrated that the molecules oscillate out-of-phase with the incident radiation at the energy of the collective mode acting as an effective \textit{metallic} layer. The physical nature of the observed mode, closely related to the collective molecule-molecule coupling, has all the attributes of a superradiant mode.

\section{Author contributions statement}
R.V. conceived the experiment(s),  C. G. and P.F. conducted the experiment(s), C. G. , P.F. and R.V. analyzed the results. The theory and numerical modeling of this work were discussed between R.V. and M.S. and was performed by M.S.

\section{Aknowledgements}
P.F. and C.G. are grateful for a Ph. D. and master fellowship, respectively, provided by the French State, managed by the French National Research Agency (ANR) in the frame of the - Investments for the future - Programme IdEx Bordeaux – LAPHIA (ANR-10-IDEX-03-02). The numerical modeling performed by M.S. is supported by the Air Force Office of Scientific Research under grant No. FA9550-15-1-0189 and Binational Science Foundation under grant No. 2014113.


\begin{thebibliography}{30}%
\makeatletter
\providecommand \@ifxundefined [1]{%
 \@ifx{#1\undefined}
}%
\providecommand \@ifnum [1]{%
 \ifnum #1\expandafter \@firstoftwo
 \else \expandafter \@secondoftwo
 \fi
}%
\providecommand \@ifx [1]{%
 \ifx #1\expandafter \@firstoftwo
 \else \expandafter \@secondoftwo
 \fi
}%
\providecommand \natexlab [1]{#1}%
\providecommand \enquote  [1]{``#1''}%
\providecommand \bibnamefont  [1]{#1}%
\providecommand \bibfnamefont [1]{#1}%
\providecommand \citenamefont [1]{#1}%
\providecommand \href@noop [0]{\@secondoftwo}%
\providecommand \href [0]{\begingroup \@sanitize@url \@href}%
\providecommand \@href[1]{\@@startlink{#1}\@@href}%
\providecommand \@@href[1]{\endgroup#1\@@endlink}%
\providecommand \@sanitize@url [0]{\catcode `\\12\catcode `\$12\catcode
  `\&12\catcode `\#12\catcode `\^12\catcode `\_12\catcode `\%12\relax}%
\providecommand \@@startlink[1]{}%
\providecommand \@@endlink[0]{}%
\providecommand \url  [0]{\begingroup\@sanitize@url \@url }%
\providecommand \@url [1]{\endgroup\@href {#1}{\urlprefix }}%
\providecommand \urlprefix  [0]{URL }%
\providecommand \Eprint [0]{\href }%
\providecommand \doibase [0]{http://dx.doi.org/}%
\providecommand \selectlanguage [0]{\@gobble}%
\providecommand \bibinfo  [0]{\@secondoftwo}%
\providecommand \bibfield  [0]{\@secondoftwo}%
\providecommand \translation [1]{[#1]}%
\providecommand \BibitemOpen [0]{}%
\providecommand \bibitemStop [0]{}%
\providecommand \bibitemNoStop [0]{.\EOS\space}%
\providecommand \EOS [0]{\spacefactor3000\relax}%
\providecommand \BibitemShut  [1]{\csname bibitem#1\endcsname}%
\let\auto@bib@innerbib\@empty
%</preamble>
\bibitem [{\citenamefont {Stockman}(2011)}]{Stockman:11}%
  \BibitemOpen
  \bibfield  {author} {\bibinfo {author} {\bibfnamefont {M.~I.}\ \bibnamefont
  {Stockman}},\ }\href {\doibase 10.1364/OE.19.022029} {\bibfield  {journal}
  {\bibinfo  {journal} {Opt. Express}\ }\textbf {\bibinfo {volume} {19}},\
  \bibinfo {pages} {22029} (\bibinfo {year} {2011})}\BibitemShut {NoStop}%
\bibitem [{\citenamefont {Murray}\ and\ \citenamefont
  {Barnes}(2007)}]{ADMA:ADMA200700678}%
  \BibitemOpen
  \bibfield  {author} {\bibinfo {author} {\bibfnamefont {W.~A.}\ \bibnamefont
  {Murray}}\ and\ \bibinfo {author} {\bibfnamefont {W.~L.}\ \bibnamefont
  {Barnes}},\ }\href {\doibase 10.1002/adma.200700678} {\bibfield  {journal}
  {\bibinfo  {journal} {Adv. Mater.}\ }\textbf {\bibinfo {volume} {19}},\
  \bibinfo {pages} {3771} (\bibinfo {year} {2007})}\BibitemShut {NoStop}%
\bibitem [{\citenamefont {Gramotnev}\ and\ \citenamefont
  {Bozhevolnyi}(2014)}]{Gramotnev:2014aa}%
  \BibitemOpen
  \bibfield  {author} {\bibinfo {author} {\bibfnamefont {D.~K.}\ \bibnamefont
  {Gramotnev}}\ and\ \bibinfo {author} {\bibfnamefont {S.~I.}\ \bibnamefont
  {Bozhevolnyi}},\ }\href {http://dx.doi.org/10.1038/nphoton.2013.232}
  {\bibfield  {journal} {\bibinfo  {journal} {Nat. Photon.}\ }\textbf {\bibinfo
  {volume} {8}},\ \bibinfo {pages} {13} (\bibinfo {year} {2014})}\BibitemShut
  {NoStop}%
\bibitem [{\citenamefont {Pelton}(2015)}]{Pelton:2015aa}%
  \BibitemOpen
  \bibfield  {author} {\bibinfo {author} {\bibfnamefont {M.}~\bibnamefont
  {Pelton}},\ }\href {http://dx.doi.org/10.1038/nphoton.2015.103} {\bibfield
  {journal} {\bibinfo  {journal} {Nat. Photon.}\ }\textbf {\bibinfo {volume}
  {9}},\ \bibinfo {pages} {427} (\bibinfo {year} {2015})}\BibitemShut {NoStop}%
\bibitem [{\citenamefont {Lodahl}\ \emph {et~al.}(2015)\citenamefont {Lodahl},
  \citenamefont {Mahmoodian},\ and\ \citenamefont {Stobbe}}]{Lodahl:2015aa}%
  \BibitemOpen
  \bibfield  {author} {\bibinfo {author} {\bibfnamefont {P.}~\bibnamefont
  {Lodahl}}, \bibinfo {author} {\bibfnamefont {S.}~\bibnamefont {Mahmoodian}},
  \ and\ \bibinfo {author} {\bibfnamefont {S.}~\bibnamefont {Stobbe}},\ }\href
  {http://link.aps.org/doi/10.1103/RevModPhys.87.347} {\bibfield  {journal}
  {\bibinfo  {journal} {Rev. Mod. Phys.}\ }\textbf {\bibinfo {volume} {87}},\
  \bibinfo {pages} {347} (\bibinfo {year} {2015})}\BibitemShut {NoStop}%
\bibitem [{\citenamefont {Bellessa}\ \emph {et~al.}(2009)\citenamefont
  {Bellessa}, \citenamefont {Symonds}, \citenamefont {Vynck}, \citenamefont
  {Lemaitre}, \citenamefont {Brioude}, \citenamefont {Beaur}, \citenamefont
  {Plenet}, \citenamefont {Viste}, \citenamefont {Felbacq}, \citenamefont
  {Cambril},\ and\ \citenamefont {Valvin}}]{Bellessa2009}%
  \BibitemOpen
  \bibfield  {author} {\bibinfo {author} {\bibfnamefont {J.}~\bibnamefont
  {Bellessa}}, \bibinfo {author} {\bibfnamefont {C.}~\bibnamefont {Symonds}},
  \bibinfo {author} {\bibfnamefont {K.}~\bibnamefont {Vynck}}, \bibinfo
  {author} {\bibfnamefont {A.}~\bibnamefont {Lemaitre}}, \bibinfo {author}
  {\bibfnamefont {A.}~\bibnamefont {Brioude}}, \bibinfo {author} {\bibfnamefont
  {L.}~\bibnamefont {Beaur}}, \bibinfo {author} {\bibfnamefont {J.~C.}\
  \bibnamefont {Plenet}}, \bibinfo {author} {\bibfnamefont {P.}~\bibnamefont
  {Viste}}, \bibinfo {author} {\bibfnamefont {D.}~\bibnamefont {Felbacq}},
  \bibinfo {author} {\bibfnamefont {E.}~\bibnamefont {Cambril}}, \ and\
  \bibinfo {author} {\bibfnamefont {P.}~\bibnamefont {Valvin}},\ }\href
  {\doibase 10.1103/PhysRevB.80.033303} {\bibfield  {journal} {\bibinfo
  {journal} {Phys. Rev. B}\ }\textbf {\bibinfo {volume} {80}},\ \bibinfo
  {pages} {033303} (\bibinfo {year} {2009})}\BibitemShut {NoStop}%
\bibitem [{\citenamefont {Balci}(2013)}]{Balci2013}%
  \BibitemOpen
  \bibfield  {author} {\bibinfo {author} {\bibfnamefont {S.}~\bibnamefont
  {Balci}},\ }\href {\doibase 10.1364/OL.38.004498} {\bibfield  {journal}
  {\bibinfo  {journal} {Opt. Lett.}\ }\textbf {\bibinfo {volume} {38}},\
  \bibinfo {pages} {4498} (\bibinfo {year} {2013})}\BibitemShut {NoStop}%
\bibitem [{\citenamefont {Zengin}\ \emph {et~al.}(2013)\citenamefont {Zengin},
  \citenamefont {Johansson}, \citenamefont {Johansson}, \citenamefont
  {Antosiewicz}, \citenamefont {K{\"{a}}ll},\ and\ \citenamefont
  {Shegai}}]{Zengin2013}%
  \BibitemOpen
  \bibfield  {author} {\bibinfo {author} {\bibfnamefont {G.}~\bibnamefont
  {Zengin}}, \bibinfo {author} {\bibfnamefont {G.}~\bibnamefont {Johansson}},
  \bibinfo {author} {\bibfnamefont {P.}~\bibnamefont {Johansson}}, \bibinfo
  {author} {\bibfnamefont {T.~J.}\ \bibnamefont {Antosiewicz}}, \bibinfo
  {author} {\bibfnamefont {M.}~\bibnamefont {K{\"{a}}ll}}, \ and\ \bibinfo
  {author} {\bibfnamefont {T.}~\bibnamefont {Shegai}},\ }\href {\doibase
  10.1038/srep03074} {\bibfield  {journal} {\bibinfo  {journal} {Sci. Rep.}\
  }\textbf {\bibinfo {volume} {3}},\ \bibinfo {pages} {3074} (\bibinfo {year}
  {2013})}\BibitemShut {NoStop}%
\bibitem [{\citenamefont {Hakami}\ \emph {et~al.}(2014)\citenamefont {Hakami},
  \citenamefont {Wang},\ and\ \citenamefont {Zubairy}}]{Hakami2014}%
  \BibitemOpen
  \bibfield  {author} {\bibinfo {author} {\bibfnamefont {J.}~\bibnamefont
  {Hakami}}, \bibinfo {author} {\bibfnamefont {L.}~\bibnamefont {Wang}}, \ and\
  \bibinfo {author} {\bibfnamefont {M.~S.}\ \bibnamefont {Zubairy}},\ }\href
  {\doibase 10.1103/PhysRevA.89.053835} {\bibfield  {journal} {\bibinfo
  {journal} {Phys. Rev. A}\ }\textbf {\bibinfo {volume} {89}},\ \bibinfo
  {pages} {053835} (\bibinfo {year} {2014})}\BibitemShut {NoStop}%
\bibitem [{\citenamefont {Bellessa}\ \emph {et~al.}(2004)\citenamefont
  {Bellessa}, \citenamefont {Bonnand}, \citenamefont {Plenet},\ and\
  \citenamefont {Mugnier}}]{Bellessa2004}%
  \BibitemOpen
  \bibfield  {author} {\bibinfo {author} {\bibfnamefont {J.}~\bibnamefont
  {Bellessa}}, \bibinfo {author} {\bibfnamefont {C.}~\bibnamefont {Bonnand}},
  \bibinfo {author} {\bibfnamefont {J.~C.}\ \bibnamefont {Plenet}}, \ and\
  \bibinfo {author} {\bibfnamefont {J.}~\bibnamefont {Mugnier}},\ }\href
  {\doibase 10.1103/PhysRevLett.93.036404} {\bibfield  {journal} {\bibinfo
  {journal} {Physical Review Letters}\ }\textbf {\bibinfo {volume} {93}},\
  \bibinfo {pages} {036404} (\bibinfo {year} {2004})}\BibitemShut {NoStop}%
\bibitem [{\citenamefont {Hakala}\ \emph {et~al.}(2009)\citenamefont {Hakala},
  \citenamefont {Toppari}, \citenamefont {Kuzyk}, \citenamefont {Pettersson},
  \citenamefont {Tikkanen}, \citenamefont {Kunttu},\ and\ \citenamefont
  {T{\"{o}}rm{\"{a}}}}]{Hakala2009}%
  \BibitemOpen
  \bibfield  {author} {\bibinfo {author} {\bibfnamefont {T.}~\bibnamefont
  {Hakala}}, \bibinfo {author} {\bibfnamefont {J.}~\bibnamefont {Toppari}},
  \bibinfo {author} {\bibfnamefont {A.}~\bibnamefont {Kuzyk}}, \bibinfo
  {author} {\bibfnamefont {M.}~\bibnamefont {Pettersson}}, \bibinfo {author}
  {\bibfnamefont {H.}~\bibnamefont {Tikkanen}}, \bibinfo {author}
  {\bibfnamefont {H.}~\bibnamefont {Kunttu}}, \ and\ \bibinfo {author}
  {\bibfnamefont {P.}~\bibnamefont {T{\"{o}}rm{\"{a}}}},\ }\href {\doibase
  10.1103/PhysRevLett.103.053602} {\bibfield  {journal} {\bibinfo  {journal}
  {Phys. Rev. Lett.}\ }\textbf {\bibinfo {volume} {103}},\ \bibinfo {pages}
  {053602} (\bibinfo {year} {2009})}\BibitemShut {NoStop}%
\bibitem [{\citenamefont {Balci}\ \emph {et~al.}(2012)\citenamefont {Balci},
  \citenamefont {Kocabas}, \citenamefont {Ates}, \citenamefont {Karademir},
  \citenamefont {Salihoglu},\ and\ \citenamefont {Aydinli}}]{Balci2012}%
  \BibitemOpen
  \bibfield  {author} {\bibinfo {author} {\bibfnamefont {S.}~\bibnamefont
  {Balci}}, \bibinfo {author} {\bibfnamefont {C.}~\bibnamefont {Kocabas}},
  \bibinfo {author} {\bibfnamefont {S.}~\bibnamefont {Ates}}, \bibinfo {author}
  {\bibfnamefont {E.}~\bibnamefont {Karademir}}, \bibinfo {author}
  {\bibfnamefont {O.}~\bibnamefont {Salihoglu}}, \ and\ \bibinfo {author}
  {\bibfnamefont {A.}~\bibnamefont {Aydinli}},\ }\href {\doibase
  10.1103/PhysRevB.86.235402} {\bibfield  {journal} {\bibinfo  {journal} {Phys.
  Rev. B - Condens. Matter Mater. Phys.}\ }\textbf {\bibinfo {volume} {86}},\
  \bibinfo {pages} {1} (\bibinfo {year} {2012})}\BibitemShut {NoStop}%
\bibitem [{\citenamefont {T{\"{o}}rm{\"{a}}}\ and\ \citenamefont
  {Barnes}(2015)}]{Torma2015a}%
  \BibitemOpen
  \bibfield  {author} {\bibinfo {author} {\bibfnamefont {P.}~\bibnamefont
  {T{\"{o}}rm{\"{a}}}}\ and\ \bibinfo {author} {\bibfnamefont {W.~L.}\
  \bibnamefont {Barnes}},\ }\href {\doibase 10.1088/0034-4885/78/1/013901}
  {\bibfield  {journal} {\bibinfo  {journal} {Rep. Prog. Phys.}\ }\textbf
  {\bibinfo {volume} {78}},\ \bibinfo {pages} {013901} (\bibinfo {year}
  {2015})},\ \Eprint {http://arxiv.org/abs/1405.1661} {arXiv:1405.1661}
  \BibitemShut {NoStop}%
\bibitem [{\citenamefont {Khitrova}\ \emph {et~al.}(2006)\citenamefont
  {Khitrova}, \citenamefont {Gibbs}, \citenamefont {Kira}, \citenamefont
  {Koch},\ and\ \citenamefont {Scherer}}]{Khitrova:2006aa}%
  \BibitemOpen
  \bibfield  {author} {\bibinfo {author} {\bibfnamefont {G.}~\bibnamefont
  {Khitrova}}, \bibinfo {author} {\bibfnamefont {H.~M.}\ \bibnamefont {Gibbs}},
  \bibinfo {author} {\bibfnamefont {M.}~\bibnamefont {Kira}}, \bibinfo {author}
  {\bibfnamefont {S.~W.}\ \bibnamefont {Koch}}, \ and\ \bibinfo {author}
  {\bibfnamefont {A.}~\bibnamefont {Scherer}},\ }\href
  {http://dx.doi.org/10.1038/nphys227} {\bibfield  {journal} {\bibinfo
  {journal} {Nat. Phys.}\ }\textbf {\bibinfo {volume} {2}},\ \bibinfo {pages}
  {81} (\bibinfo {year} {2006})}\BibitemShut {NoStop}%
\bibitem [{\citenamefont {Schlather}\ \emph {et~al.}(2013)\citenamefont
  {Schlather}, \citenamefont {Large}, \citenamefont {Urban}, \citenamefont
  {Nordlander},\ and\ \citenamefont {Halas}}]{doi:10.1021/nl4014887}%
  \BibitemOpen
  \bibfield  {author} {\bibinfo {author} {\bibfnamefont {A.~E.}\ \bibnamefont
  {Schlather}}, \bibinfo {author} {\bibfnamefont {N.}~\bibnamefont {Large}},
  \bibinfo {author} {\bibfnamefont {A.~S.}\ \bibnamefont {Urban}}, \bibinfo
  {author} {\bibfnamefont {P.}~\bibnamefont {Nordlander}}, \ and\ \bibinfo
  {author} {\bibfnamefont {N.~J.}\ \bibnamefont {Halas}},\ }\href {\doibase
  10.1021/nl4014887} {\bibfield  {journal} {\bibinfo  {journal} {Nano Lett.}\
  }\textbf {\bibinfo {volume} {13}},\ \bibinfo {pages} {3281} (\bibinfo {year}
  {2013})}\BibitemShut {NoStop}%
\bibitem [{\citenamefont {Salomon}\ \emph {et~al.}(2012)\citenamefont
  {Salomon}, \citenamefont {Gordon}, \citenamefont {Prior}, \citenamefont
  {Seideman},\ and\ \citenamefont {Sukharev}}]{PhysRevLett.109.073002}%
  \BibitemOpen
  \bibfield  {author} {\bibinfo {author} {\bibfnamefont {A.}~\bibnamefont
  {Salomon}}, \bibinfo {author} {\bibfnamefont {R.~J.}\ \bibnamefont {Gordon}},
  \bibinfo {author} {\bibfnamefont {Y.}~\bibnamefont {Prior}}, \bibinfo
  {author} {\bibfnamefont {T.}~\bibnamefont {Seideman}}, \ and\ \bibinfo
  {author} {\bibfnamefont {M.}~\bibnamefont {Sukharev}},\ }\href {\doibase
  10.1103/PhysRevLett.109.073002} {\bibfield  {journal} {\bibinfo  {journal}
  {Phys. Rev. Lett.}\ }\textbf {\bibinfo {volume} {109}},\ \bibinfo {pages}
  {073002} (\bibinfo {year} {2012})}\BibitemShut {NoStop}%
\bibitem [{\citenamefont {Ungureanu}\ \emph {et~al.}(2013)\citenamefont
  {Ungureanu}, \citenamefont {Kolaric}, \citenamefont {Chen}, \citenamefont
  {Hillenbrand},\ and\ \citenamefont {Vall{\'{e}}e}}]{Ungureanu2013}%
  \BibitemOpen
  \bibfield  {author} {\bibinfo {author} {\bibfnamefont {S.}~\bibnamefont
  {Ungureanu}}, \bibinfo {author} {\bibfnamefont {B.}~\bibnamefont {Kolaric}},
  \bibinfo {author} {\bibfnamefont {J.}~\bibnamefont {Chen}}, \bibinfo {author}
  {\bibfnamefont {R.}~\bibnamefont {Hillenbrand}}, \ and\ \bibinfo {author}
  {\bibfnamefont {R.~R. a.~L.}\ \bibnamefont {Vall{\'{e}}e}},\ }\href {\doibase
  10.1515/nanoph-2013-0004} {\bibfield  {journal} {\bibinfo  {journal}
  {Nanophotonics}\ }\textbf {\bibinfo {volume} {2}},\ \bibinfo {pages} {173}
  (\bibinfo {year} {2013})}\BibitemShut {NoStop}%
\bibitem [{\citenamefont {Fauch\'{e}}\ \emph {et~al.}(2014)\citenamefont
  {Fauch\'{e}}, \citenamefont {Ungureanu}, \citenamefont {Kolaric},\ and\
  \citenamefont {Vall\'{e}e}}]{Fauche2014}%
  \BibitemOpen
  \bibfield  {author} {\bibinfo {author} {\bibfnamefont {P.}~\bibnamefont
  {Fauch\'{e}}}, \bibinfo {author} {\bibfnamefont {S.}~\bibnamefont
  {Ungureanu}}, \bibinfo {author} {\bibfnamefont {B.}~\bibnamefont {Kolaric}},
  \ and\ \bibinfo {author} {\bibfnamefont {R.~A.~L.}\ \bibnamefont
  {Vall\'{e}e}},\ }\href {\doibase 10.1039/C4TC01787K} {\bibfield  {journal}
  {\bibinfo  {journal} {J. Mater. Chem. C}\ }\textbf {\bibinfo {volume} {2}},\
  \bibinfo {pages} {10362} (\bibinfo {year} {2014})}\BibitemShut {NoStop}%
\bibitem [{\citenamefont {Landstr\"{o}m}\ \emph
  {et~al.}(2006{\natexlab{a}})\citenamefont {Landstr\"{o}m}, \citenamefont
  {Brodoceanu}, \citenamefont {Piglmayer},\ and\ \citenamefont
  {B\"{a}uerle}}]{Landstrom2006a}%
  \BibitemOpen
  \bibfield  {author} {\bibinfo {author} {\bibfnamefont {L.}~\bibnamefont
  {Landstr\"{o}m}}, \bibinfo {author} {\bibfnamefont {D.}~\bibnamefont
  {Brodoceanu}}, \bibinfo {author} {\bibfnamefont {K.}~\bibnamefont
  {Piglmayer}}, \ and\ \bibinfo {author} {\bibfnamefont {D.}~\bibnamefont
  {B\"{a}uerle}},\ }\href {\doibase 10.1007/s00339-006-3635-8} {\bibfield
  {journal} {\bibinfo  {journal} {Applied Physics A}\ }\textbf {\bibinfo
  {volume} {84}},\ \bibinfo {pages} {373} (\bibinfo {year}
  {2006}{\natexlab{a}})}\BibitemShut {NoStop}%
\bibitem [{\citenamefont {Romanov}\ \emph {et~al.}(2011)\citenamefont
  {Romanov}, \citenamefont {Korovin}, \citenamefont {Regensburger},\ and\
  \citenamefont {Peschel}}]{Romanov2011}%
  \BibitemOpen
  \bibfield  {author} {\bibinfo {author} {\bibfnamefont {S.~G.}\ \bibnamefont
  {Romanov}}, \bibinfo {author} {\bibfnamefont {A.~V.}\ \bibnamefont
  {Korovin}}, \bibinfo {author} {\bibfnamefont {A.}~\bibnamefont
  {Regensburger}}, \ and\ \bibinfo {author} {\bibfnamefont {U.}~\bibnamefont
  {Peschel}},\ }\href@noop {} {\enquote {\bibinfo {title} {{Hybrid colloidal
  plasmonic-photonic crystals}},}\ } (\bibinfo {year} {2011})\BibitemShut
  {NoStop}%
\bibitem [{\citenamefont {Landstr\"{o}m}\ \emph
  {et~al.}(2006{\natexlab{b}})\citenamefont {Landstr\"{o}m}, \citenamefont
  {Arnold}, \citenamefont {Brodoceanu}, \citenamefont {Piglmayer},\ and\
  \citenamefont {B\"{a}uerle}}]{Landstrom2006}%
  \BibitemOpen
  \bibfield  {author} {\bibinfo {author} {\bibfnamefont {L.}~\bibnamefont
  {Landstr\"{o}m}}, \bibinfo {author} {\bibfnamefont {N.}~\bibnamefont
  {Arnold}}, \bibinfo {author} {\bibfnamefont {D.}~\bibnamefont {Brodoceanu}},
  \bibinfo {author} {\bibfnamefont {K.}~\bibnamefont {Piglmayer}}, \ and\
  \bibinfo {author} {\bibfnamefont {D.}~\bibnamefont {B\"{a}uerle}},\ }\href
  {\doibase 10.1007/s00339-006-3490-7} {\bibfield  {journal} {\bibinfo
  {journal} {Applied Physics A}\ }\textbf {\bibinfo {volume} {83}},\ \bibinfo
  {pages} {271} (\bibinfo {year} {2006}{\natexlab{b}})}\BibitemShut {NoStop}%
\bibitem [{\citenamefont {Ding}\ \emph {et~al.}(2013)\citenamefont {Ding},
  \citenamefont {Hrelescu}, \citenamefont {Arnold}, \citenamefont {Isic},\ and\
  \citenamefont {Klar}}]{Ding2013}%
  \BibitemOpen
  \bibfield  {author} {\bibinfo {author} {\bibfnamefont {B.}~\bibnamefont
  {Ding}}, \bibinfo {author} {\bibfnamefont {C.}~\bibnamefont {Hrelescu}},
  \bibinfo {author} {\bibfnamefont {N.}~\bibnamefont {Arnold}}, \bibinfo
  {author} {\bibfnamefont {G.}~\bibnamefont {Isic}}, \ and\ \bibinfo {author}
  {\bibfnamefont {T.~A.}\ \bibnamefont {Klar}},\ }\href {\doibase
  10.1021/nl3035114} {\bibfield  {journal} {\bibinfo  {journal} {Nano letters}\
  }\textbf {\bibinfo {volume} {13}},\ \bibinfo {pages} {378} (\bibinfo {year}
  {2013})}\BibitemShut {NoStop}%
\bibitem [{\citenamefont {Landstr\"{o}m}\ \emph {et~al.}(2005)\citenamefont
  {Landstr\"{o}m}, \citenamefont {Brodoceanu}, \citenamefont {Arnold},
  \citenamefont {Piglmayer},\ and\ \citenamefont
  {B\"{a}uerle}}]{Landstrom2005}%
  \BibitemOpen
  \bibfield  {author} {\bibinfo {author} {\bibfnamefont {L.}~\bibnamefont
  {Landstr\"{o}m}}, \bibinfo {author} {\bibfnamefont {D.}~\bibnamefont
  {Brodoceanu}}, \bibinfo {author} {\bibfnamefont {N.}~\bibnamefont {Arnold}},
  \bibinfo {author} {\bibfnamefont {K.}~\bibnamefont {Piglmayer}}, \ and\
  \bibinfo {author} {\bibfnamefont {D.}~\bibnamefont {B\"{a}uerle}},\ }\href
  {\doibase 10.1007/s00339-005-3309-y} {\bibfield  {journal} {\bibinfo
  {journal} {Applied Physics A}\ }\textbf {\bibinfo {volume} {81}},\ \bibinfo
  {pages} {911} (\bibinfo {year} {2005})}\BibitemShut {NoStop}%
\bibitem [{\citenamefont {Ma}\ \emph {et~al.}(2003)\citenamefont {Ma},
  \citenamefont {Lu}, \citenamefont {Brock}, \citenamefont {Jacobs},
  \citenamefont {Yang},\ and\ \citenamefont {Hu}}]{Ma2003}%
  \BibitemOpen
  \bibfield  {author} {\bibinfo {author} {\bibfnamefont {X.}~\bibnamefont
  {Ma}}, \bibinfo {author} {\bibfnamefont {J.~Q.}\ \bibnamefont {Lu}}, \bibinfo
  {author} {\bibfnamefont {R.~S.}\ \bibnamefont {Brock}}, \bibinfo {author}
  {\bibfnamefont {K.~M.}\ \bibnamefont {Jacobs}}, \bibinfo {author}
  {\bibfnamefont {P.}~\bibnamefont {Yang}}, \ and\ \bibinfo {author}
  {\bibfnamefont {X.-H.}\ \bibnamefont {Hu}},\ }\href {\doibase
  10.1088/0031-9155/48/24/013} {\bibfield  {journal} {\bibinfo  {journal}
  {Physics in Medicine and Biology}\ }\textbf {\bibinfo {volume} {48}},\
  \bibinfo {pages} {4165} (\bibinfo {year} {2003})}\BibitemShut {NoStop}%
\bibitem [{\citenamefont {Johnson}\ and\ \citenamefont
  {Christy}(1972)}]{Johnson1972}%
  \BibitemOpen
  \bibfield  {author} {\bibinfo {author} {\bibfnamefont {P.~B.}\ \bibnamefont
  {Johnson}}\ and\ \bibinfo {author} {\bibfnamefont {R.~W.}\ \bibnamefont
  {Christy}},\ }\href {\doibase 10.1103/PhysRevB.6.4370} {\bibfield  {journal}
  {\bibinfo  {journal} {Physical Review B}\ }\textbf {\bibinfo {volume} {6}},\
  \bibinfo {pages} {4370} (\bibinfo {year} {1972})}\BibitemShut {NoStop}%
\bibitem [{\citenamefont {Agranovich}\ \emph {et~al.}(2003)\citenamefont
  {Agranovich}, \citenamefont {Litinskaia},\ and\ \citenamefont
  {Lidzey}}]{Agranovich2003}%
  \BibitemOpen
  \bibfield  {author} {\bibinfo {author} {\bibfnamefont {V.~M.}\ \bibnamefont
  {Agranovich}}, \bibinfo {author} {\bibfnamefont {M.}~\bibnamefont
  {Litinskaia}}, \ and\ \bibinfo {author} {\bibfnamefont {D.~G.}\ \bibnamefont
  {Lidzey}},\ }\href {\doibase 10.1103/PhysRevB.67.085311} {\bibfield
  {journal} {\bibinfo  {journal} {Phys. Rev. B}\ }\textbf {\bibinfo {volume}
  {67}},\ \bibinfo {pages} {085311} (\bibinfo {year} {2003})}\BibitemShut
  {NoStop}%
\bibitem [{\citenamefont {Sukharev}\ and\ \citenamefont
  {Nitzan}(2011)}]{PhysRevA.84.043802}%
  \BibitemOpen
  \bibfield  {author} {\bibinfo {author} {\bibfnamefont {M.}~\bibnamefont
  {Sukharev}}\ and\ \bibinfo {author} {\bibfnamefont {A.}~\bibnamefont
  {Nitzan}},\ }\href {\doibase 10.1103/PhysRevA.84.043802} {\bibfield
  {journal} {\bibinfo  {journal} {Phys. Rev. A}\ }\textbf {\bibinfo {volume}
  {84}},\ \bibinfo {pages} {043802} (\bibinfo {year} {2011})}\BibitemShut
  {NoStop}%
\bibitem [{\citenamefont {Blake}\ and\ \citenamefont
  {Sukharev}(2015)}]{Blake:2015aa}%
  \BibitemOpen
  \bibfield  {author} {\bibinfo {author} {\bibfnamefont {A.}~\bibnamefont
  {Blake}}\ and\ \bibinfo {author} {\bibfnamefont {M.}~\bibnamefont
  {Sukharev}},\ }\href {http://link.aps.org/doi/10.1103/PhysRevB.92.035433}
  {\bibfield  {journal} {\bibinfo  {journal} {Phys. Rev. B}\ }\textbf {\bibinfo
  {volume} {92}},\ \bibinfo {pages} {035433} (\bibinfo {year}
  {2015})}\BibitemShut {NoStop}%
\bibitem [{\citenamefont {Sukharev}\ and\ \citenamefont
  {Nitzan}(2016)}]{Sukharev:2016aa}%
  \BibitemOpen
  \bibfield  {author} {\bibinfo {author} {\bibfnamefont {M.}~\bibnamefont
  {Sukharev}}\ and\ \bibinfo {author} {\bibfnamefont {A.}~\bibnamefont
  {Nitzan}},\ }\href
  {http://scitation.aip.org/content/aip/journal/jcp/144/14/10.1063/1.4945446}
  {\bibfield  {journal} {\bibinfo  {journal} {The Journal of Chemical Physics}\
  }\textbf {\bibinfo {volume} {144}},\ \bibinfo {pages} {144703} (\bibinfo
  {year} {2016})}\BibitemShut {NoStop}%
\bibitem [{\citenamefont {Antosiewicz}\ \emph {et~al.}(2014)\citenamefont
  {Antosiewicz}, \citenamefont {Apell},\ and\ \citenamefont
  {Shegai}}]{doi:10.1021/ph500032d}%
  \BibitemOpen
  \bibfield  {author} {\bibinfo {author} {\bibfnamefont {T.~J.}\ \bibnamefont
  {Antosiewicz}}, \bibinfo {author} {\bibfnamefont {S.~P.}\ \bibnamefont
  {Apell}}, \ and\ \bibinfo {author} {\bibfnamefont {T.}~\bibnamefont
  {Shegai}},\ }\href {\doibase 10.1021/ph500032d} {\bibfield  {journal}
  {\bibinfo  {journal} {ACS Photonics}\ }\textbf {\bibinfo {volume} {1}},\
  \bibinfo {pages} {454} (\bibinfo {year} {2014})}\BibitemShut {NoStop}%
\end{thebibliography}
\end{document}